\documentstyle[12pt,aaspp4]{article}
\begin{document}
\renewcommand\floatpagefraction{.9}
\renewcommand\topfraction{1}
\renewcommand\bottomfraction{1}
\renewcommand\floatsep{12pt}
\renewcommand\textfraction{0}
\renewcommand\intextsep{12pt}

\title{ THE HOST GALAXIES OF THREE RADIO-LOUD QUASARS: 3C~48, 3C~345,
AND B2~1425+267
\footnote{Based on observations with the NASA/ESA Hubble Space
Telescope, obtained at the Space Telescope Science Institute, which is
operated by the Association of Universities for Research in Astronomy,
Inc., under NASA contract NAS5-26555.\smallskip}}
\author{Sofia Kirhakos and John N. Bahcall}
\affil{Institute for Advanced Study, School of Natural Sciences,
Princeton, NJ~08540}
\author{Donald P. Schneider}
\affil{Department of Astronomy and Astrophysics, The Pennsylvania 
State University, University Park, PA 16802}
\centerline{and}
\author{Jerome Kristian
\footnote{Deceased}}
\affil{Observatories of the Carnegie Institution of Washington,
Pasadena, CA 91101}
\begin{abstract}

Observations with the Wide-Field/Planetary Camera-2 of the {\it Hubble
Space Telescope (HST)} are presented for three radio-loud quasars:
3C~48 ($z=0.367$), B2~1425+267 ($z=0.366$), and 3C~345 ($z=0.594$).
All three quasars have luminous ($\sim 4\times L^*$) galaxies as
hosts, which are either elliptical (B2~1425+267 and 3C~345) or
interacting (3C~48), and all hosts are $0.5 - 1.0$~mag bluer in ($V-I$) than
other galaxies with the same overall morphology at similar redshifts
to the quasars. The host of 3C~48 has many H~II regions and a very
extended tidal tail.

All nine of the radio-loud quasars studied here and in Bahcall
et~al. (1997) either have bright elliptical hosts or occur in
interacting systems.  There is a robust correlation between the radio
emission of the quasar and the luminosity of host galaxy; the
radio-loud quasars reside in galaxies that are on average $\sim 1$~mag
brighter than  hosts of the radio-quiet quasars.

\end{abstract}

\section{INTRODUCTION}
\label{introduction}

The characteristics of the host galaxies in which quasars reside may
provide clues about the mechanisms that produce the observed energetic
phenomena. To understand quasars, we would like to know answers to the
following key questions: What kind of galaxies display the quasar
phenomena?  How old are the galaxies? Do the hosts show signs of
interactions or of star formation? How are the quasars fed? How long
is a quasar active in any galaxy? What fraction of galaxies show
quasar phenomena?

The search for quasar host galaxies to help answer these questions was
initiated by the pioneering work, using photographic plates, of
Kristian~(1973) and Wyckoff et~al. (1980). The use of two-dimensional
electronic detectors and Charge Coupled Devices (CCDs) made possible
major improvements in the studies of quasar environments; some
representative papers are Hutchings, Janson, \& Neff~(1989), Hutchings
\& Neff (1992), Dunlop et~al.~(1993), and McLeod \& Rieke (1994a,b).
These ground-based studies indicated the presence of host galaxies of
many quasars and suggested that some of the hosts appeared to be
disturbed by a gravitational encounter ({\it e.g.}, Hutchings \&
Campbell 1983, Hutchings, Crampton, \& Campbell 1984).  Several
spectroscopic studies of the nebulosity surrounding quasars have
indicated that the nature of the emission is starlight from the
underlying galaxy (Boroson, Oke, \& Green 1982; Boroson \& Oke 1982,
1984; Balick \& Heckman 1983, Kukula et~al. 1996). In particular,
Kukula et~al. obtained optical off-nuclear
spectra of active galaxies and estimated the ages of the stellar
populations using spectral synthesis models.

The $HST$ opened a new window to the study of quasar host
galaxies. The high angular resolution of the $HST$ has produced
spectacular images of a number of quasars (Bahcall et~al. 1994,
1995a,b, 1996a,b, 1997; Hutchings et~al. 1994; Disney et~al. 1995;
Hutchings \& Morris 1995; Boyce et~al. 1996; Hooper et~al. 1997).  The
$HST$ images also revealed the presence of surprisingly close
companions; for example, PKS~2349+014 (Bahcall et~al. 1995a) has a
companion galaxy located only 3.5~kpc in projection (for $H_0~=~100~
{\rm km~s^{-1}~Mpc^{-1}}$ and $\Omega_0~=~1.0$) from the quasar
center.

In Bahcall et~al. (1996b, 1997), we reported the results of our study
of 20 nearby ($z < 0.30$) luminous quasars observed with the
$HST$-WFPC2 in F606W.  The quasars in this Local Sample occur in
diverse environments that include bright ellipticals, apparently
normal elliptical and spiral galaxies, very faint hosts, and complex
systems of gravitationally interacting components.  The
hosts of all six radio-loud quasars (RLQs) in the sample are bright
elliptical galaxies or interacting systems.  However, contrary to
widespread expectations based on ground-based observations of radio
galaxies and Seyfert galaxies (see Balick \& Heckman 1982 and
references therein), the radio-quiet quasars (RQQs) occur in both
elliptical {\it and } spiral galaxies.  A tendency for
high-luminosity quasar to occur in early-type galaxies was suggested
by McLeod \& Rieke (1995b).
A similar result was obtained
by Taylor et~al. (1996) based on ground-based near-infrared imaging.
Only six RLQs are included in the Local Sample, which motivates us to
obtain observations of more RLQs to better understand the connection
between the radio emission and the quasar environment.

In this paper, we present $HST$-WFPC2 observations of three additional
RLQs: 3C~48, B2~1425+267, and 3C~345. Their redshifts
are 0.367, 0.366, and 0.594, respectively, higher than all the
redshifts of the Local Sample quasars and comparable to the objects
studied by Hooper et~al. (1997). The optical luminosities of the three
quasars are in the range of luminosities covered by our Local Sample,
but are on average significantly more luminous than the Hooper et~al. (1997) sample.  3C~48 is a compact steep-spectrum radio source, with
strong asymmetric emission with respect to the core and a
sub-arcsecond radio jet (Fanti et~al. 1995; Simon et~al. 1990).
B2~1425+267 is a double-lobed radio source, with angular size of
240\arcsec\ (Rogola et~al. 1986).  3C~345 is a superluminal source,
with an unresolved flat radio spectrum and a radio jet extending out
to several arcseconds (Kollgaard, Wardle, \& Roberts 1989).

This paper is organized as follows. In \S~\ref{obssection} we describe
the observations; in \S~\ref{anasection} we describe the subtraction
of the Point-Spread
Function (PSF) and the methods of analyzing of the data; in
\S~\ref{secsize} we report on the magnitudes and colors measured for the host
galaxies; in \S\S~\ref{3c48} to \ref{3c345} we discuss individually the hosts of
3C~48, B2~1425+267, and 3C~345, respectively; in \S~\ref{companion} we
discuss the presence of companion galaxies; and in \S~\ref{discuss}
we summarize and discuss our results.
Through this paper we assume that 	
$H_0~=~100~ {\rm km~s^{-1}~Mpc^{-1}}$ and $\Omega_0~=~1.0$, unless
noted otherwise. When we quote distances and luminosities from papers
that used  different cosmological parameters, we have transformed the
values to the $H_0~=~100~ {\rm km~s^{-1}~Mpc^{-1}}$ and
$\Omega_0~=~1.0$ system.

\section{OBSERVATIONS}
\label{obssection}

The quasars 3C~48, B2~1425+267, and 3C~345 were observed with the Wide
Field/Planetary Camera-2 (WFPC2; see Biretta et~al. 1996) through the
F555W and F814W filters, which correspond approximately to the $V$ and
$I$ band-passes, respectively. 
Redshifted
[O~II]$\lambda$~3727 and [O~III]$\lambda$~4363 are included in the
F555W bandpass of the three quasars; for 3C~345, redshifted H$\beta$
and [O~III]$\lambda \lambda$ 4959,5007 are also included in the F814W
band.  B2~1425+267 was placed near the center of the Wide-Field Camera
CCD~3 (WF3), while 3C~48 and 3C~345 were placed near the center of the
Planetary Camera (PC). 
Four images of each quasar were
obtained in each filter: two short exposures (50~s) to investigate the
small scale structure close to the nucleus, and two long exposures
(1400~s in F555W and 1700~s in F814W) to examine the parent galaxy,
nearby galaxies, and other extended structure.

The image scales of the PC and WF3 are 0\farcs0455\ pixel$^{-1}$ and
0\farcs0996\ pixel$^{-1}$, respectively.  The adopted photometric
zero-point for 1~${\rm electron~sec^{-1}}$ is 24.58~mag for the F555W
filter, and 24.13~mag for the F814W band (Biretta et~al. 1996).

Narrow-band images, centered at redshifted [O~III]$\lambda$ 5007
emission, were obtained of 3C~48 using the linear ramp filter
FR680N18.  The FR680N18 filter covers the wave-band from 6800 \AA\ to
6921 \AA\ (rest wavelengths of 4974~\AA\ to 5063~\AA).  The linear
ramp filters are divided into four parallel strips where the central
wavelength across each strip varies by $\sim 6\%$. Each CCD pixel is
mapped to a unique central wavelength with a FWHM bandwidth of
approximately 1.3\% of the central wavelength (Biretta et~al. 1996).
Three exposures of 1100~s, 1300~s, and 1100~s were obtained.

The initial data processing (bias frame removal and flat-field
calibration) of the broad-band images was performed at the Space
Telescope Science Institute with their standard software package.  The
narrow-band images were not flat-fielded in the STScI calibration
pipeline; we obtained the appropriate flat-field reference image from
the $HST$ archive and flattened the data.  Cosmic rays were identified
by a pixel-by-pixel comparison of pairs of images of similar exposure
time; the intensity of a pixel containing a cosmic ray was replaced by
the scaled value of the intensity of the pixel in the other image.
The calibration of the [O~III] image was done using the Linear Ramp
Filter Calculator, and following standard procedures described in the
STScI-WFPC2 world-wide-web home page.

Figure~\ref{figqso} shows the three RLQs in the F555W and F814W bands.
The host galaxies of 3C~48 and 3C~345 are clearly visible in the F814W
images; the host of B2~1425+267 is less prominent, but can be seen
with confidence upon careful inspection of the F814W image.
A journal of the observations is given in Table~\ref{journal}.

\section{DATA ANALYSIS}
\label{anasection}

The most challenging aspect of the data analysis is removal of the
signal due to the quasar nucleus, which is a heavily saturated point
source in the long exposures.  We performed Point-Spread-Function
(PSF) subtraction and subsequent analysis using the techniques
described in Bahcall et~al. (1997).  Briefly, we found a ``best-fit''
subtraction for each image by performing a $\chi ^2$ minimization of
the residuals in an annulus between $1-3$\arcsec\ away from the quasar
center.  We used empirical (WFPC2 images of stars) and model PSFs
constructed using the Tiny~Tim software (Krist 1993), which resulted
in host galaxy magnitudes that agreed within
$0.2\pm0.1$~mag. Subsequently, all analysis was performed using the
Tiny~Tim subtracted images. Figure~\ref{figqsosub} shows the three
quasars after subtraction; galaxy parameters derived from these images
are given in Table~\ref{size}.

We determined host galaxy magnitudes by analyzing the PSF-subtracted
images using the three techniques detailed in Bahcall et~al. (1997):
aperture photometry outside of the nucleus; one-dimensional fits of
disk and de Vaucouleurs profiles to azimuthally averaged radial
profiles; and two-dimensional fits of disk and de~Vaucouleurs laws to
the images themselves. The parameters and results of these analyses
are given in Table~\ref{magtab}. The 1-D and 2-D fits are in excellent
agreement. Both approaches yield somewhat brighter magnitudes than the
aperture photometry, as expected because the annular apertures omit
light from the bright central parts of the hosts.

To check for a possible effect of the PSF-subtraction on the color of
the host galaxies we also performed two-dimensional fits on the
unsubtracted quasar images: the amplitude of the point-source was fit
simultaneously with the galaxy. The root-mean-square ($rms$)
difference between the host galaxies magnitudes obtained with the
two-dimensional fits to the unsubtracted and to the PSF-subtracted
quasar images is less than 0.3~mag; the $rms$ difference in color
$m_{\rm F555W} - m_{\rm F814W}$ is 0.1~mag.

\section{MAGNITUDES AND COLORS OF HOST GALAXIES}
\label{secsize}

Figure~\ref{figgalcolors} compares the measured colors,
 $m_{\rm F555W} - m_{\rm F814W}$, for the three hosts with the
predicted galaxy colors for different morphological types.  For the
normal galaxy colors, we use the results of Fukugita et~al. (1995).

All the three host galaxies have $m_{\rm
F555W} - m_{\rm F814W}$ colors at least 1~mag bluer than predicted for
elliptical hosts at the same redshift.

Do the  bluer colors indicate
that the light from the quasar is contaminating the measurements? To
investigate this possibility, we measured the $m_{\rm F555W} - m_{\rm
F814W}$ color in a large annular aperture centered on the quasar. The
large aperture colors average only 0.13~mag redder than the
two-dimensional fit; $m_{\rm F555W} - m_{\rm F814W}$ =
+0.7, +1.0, and +1.3 for 3C~48, B2~1425+267, and 3C~345, respectively,
for annular aperture radii 1\farcs4 to 5\farcs5, 3\farcs0 to 4\farcs0, and
1\farcs4 to 2\farcs3.  We conclude that the bluer colors of
the host galaxies are not the result of contamination by quasar light. In
plotting the colors of the host galaxies shown in
Figure~\ref{figgalcolors}, we used the average of the large annular
aperture and the two dimensional fit photometry as the
``best-estimate'', and one-half the difference between the two methods
as a measure of the uncertainty.

The measured colors suggest that the host galaxies of the three
RLQs contain a younger stellar population than expected
for elliptical or spiral galaxies at the redshifts of the quasars.

The estimated absolute $V$ and $I$ magnitudes of the quasar hosts are
included in Table~\ref{magtab}; the measured F555W and F814W
magnitudes were transformed to $V$ and $I$ by applying $k$-corrections
calculated by Fukugita et~al. (1995).  Figure~\ref{figgalcolors} shows
that the $m_{\rm F555W} - m_{\rm F814W}$ color indices for the hosts
of 3C~48 and B2~1425+267 are close to the predicted values for
irregular galaxies at redshifts similar to the quasars, while for the
host of 3C~345, the measured color index is between that predicted for
irregulars and Scd galaxies.  Given the redshift of the quasars and
the morphological type inferred from Figure~\ref{figgalcolors}, we
used Tables~6 to 8 from Fukugita et~al. (1995) to obtain the ($m_{\rm
F555W} - V$) and ($m_{\rm F814W} - I$) colors, and their Figure~10 to
obtain the $k$-correction in $V$ and $I$.

\section{3C~48 (0134$+$32)}
\label{3c48}

The radio loud-source 3C~48 was the first quasar to be optically
identified (Matthews et~al. 1961). 
The first deep images of 3C~48 revealed extended emission,
and suggested that quasars were associated with the centers of
galaxies.  3C~48 was also the first quasar to be shown to have a
spatially resolved emission-line region (Wampler et~al. 1975).

\subsection{Broad Band and [O~III] Images}
\label{sub3c48broad}

The WFPC2 broad band images, Figures~\ref{figqso}, \ref{figqsosub},
and \ref{fig3c48knots}, show that 3C~48 is surrounded by extensive
nebulosity, permeated with brighter spots, probably H~II regions, and
traces of faint tidal tails that extend tens of kiloparsecs. The
morphology of the host of 3C~48  indicates that strong gravitational
interaction has taken place, confirming the evidence of merging
suggested by previous studies (see Stockton \& Ridgway 1991 and
references therein). The F555W and F814W images look similar to one
another, although the tidal tails and extended nebulosity are more
prominent in the F814W image. The F814W image of 3C~48 was 
published by Canalizo, Stockton \& Roth (1998) when they reported the
discovery of a broad-absorption line quasar at $z=2.169$ located
41\arcsec\ southwest of 3C~48.

The tidal tail extends for $\sim 35$ kpc at PA $\sim 330^\circ$ (we
adopt the standard convention for position angle: north is at ${\rm
PA=0^\circ}$ and east at ${\rm PA=90^\circ}$); it then curves almost
in a right angle to the southwest for another $\sim 20$ kpc.  The
surface brightness of the tidal tail in the F814W images ranges from
23.2 mag~arcsec$^{2}$ at $\sim$ 4\arcsec\ (12 kpc) from the quasar
center to $\sim 25$ mag~arcsec$^{2}$ at the outer parts. The
faint-nebulosity tail can be seen in the F814W image in
Figure~\ref{fig3c48knots} (upper-right potion of the figure).

The WFPC2 broad-band images reveal a bright region centered at $\sim
1\farcs2$ northeast from the center of the quasar (see
Figure~\ref{figqsosub}).  The morphology of this feature is
complex; on top of the higher-surface-brightness region there is a
ring (indicated by an arrow in the F555W image in
Figure~\ref{figqsosub}). This ring-feature resembles ghost images seen
in previous WFPC2 data (PG~1307+085, Bahcall et~al. 1997), so the ring may be  an
optical artifact.  The morphology of the northeast bright emission
region is similar in both F555W and F814W images.

Figure~\ref{fig3c48knots} compares  the broad-band
and the [O~III]$\lambda 5007$~\AA\ images of 3C~48, in which the brightest
knots are marked for identification.  The upper panels show the F555W
(left) and F814W (right) images with a linear scale that matches that
of the [O~III] image (lower panels). 
The lower-right panel of Figure~\ref{fig3c48knots} shows the [O~III]
image with a high contrast stretch to maximize the visibility of the
bright H~II region at 0\farcs8 south of the quasar.  We do not detect
a counterpart for this emitting region in the  F555W and F814W
images.  There are five knots in common between the [O~III] and the
broad-band images: c, i, l, o, q.

\subsection{Comparison with Previous Studies}
\label{sub3c48previous}

Boroson \& Oke (1982) obtained spectra of the faint nebulosity
2\arcsec\ north and south of the center of 3C~48.  They detected
narrow emission lines ([O~II]$\lambda$~3727~\AA, [O~III]$\lambda
\lambda$ 4959, 5007~\AA, and [O~I]$\lambda$ 6300~\AA) plus a continuum
dominated by hot stars at both positions, suggesting that the host is
a spiral galaxy.  Stellar absorption lines were also detected in the
nebulosity; the integrated light has both a $B-V$ color and absorption
features indicative of middle to late A stars, suggesting that a
massive star formation burst was triggered at $\sim 10^8$ years ago.
Boroson \& Oke (1982) estimated that the magnitude of the host at
5500~\AA\ is $\sim 18.5$~mag. This is in qualitative agreement with
the magnitude we estimated with the one-dimensional model fit
(17.9~mag).  

Malkan (1984) reports a blue $v-g$ color for the 3C~48 host,
consistent with Boroson and Oke's spectroscopy. Malkan suggests that
the host of 3C~48 is an elliptical galaxy which has brightened as a
result of a burst of star formation; fitting a de Vaucouleurs model to
the luminosity profile, Malkan estimates the host galaxy to have $M_B
= -22.4$ and $r_{\rm eff} = 16$~kpc.  Applying the Fukugita
et~al. (1995) color transformation relations, the host galaxy measured
by Malkan is $\sim 0.7$~mag brighter in $V$ than our measurements, and
the effective radius quoted by Malkan is a factor $\sim 2$ larger than
the value given by our two-dimensional model (see Table~\ref{magtab}).

Stockton \& Ridgway (1991) obtained a narrow-band image of 3C~48
centered on the [O~III]$\lambda 5007$\ line and three line-free band
images centered at 0.58, 0.77, and 2.1 $\micron$.  In the continuum
images, Stockton \& Ridgway detected a high-surface-brightness region
centered at $\sim 1$ \arcsec\ to the northeast of the quasar nucleus,
which they suggested might be the nucleus of a galaxy in the process
of merging with the host galaxy of 3C~48 (they designated this region
3C~48A).  We detected a bright emission region at approximately the
same position as 3C~48A in the F555W and F814W images.  In our
broad-band $HST$ images the morphology of this feature is complex and
might be contaminated by a ghost image (see arrow in
Figure~\ref{figqsosub}).  We do not detect an obvious secondary nucleus
in this region.

In their [O~III] image, Stockton \& Ridgway detected diffuse emission
surrounding the quasar, and some bright knots, mostly to the north and
west.  The brightest [O~III] emitting region is $\sim 3\farcs8$ north
to the nucleus, corresponding to the bright knots l and o in our
Figure~\ref{fig3c48knots} .  They pointed out that the presence of this
bright [O~III] emission region to the north of the quasar explained
why many authors found a strong asymmetry in the detection of emission
in spectroscopic observations centered $\sim 4$\arcsec\ north and
south of the quasar.

The radio structure of 3C~48 at 329 MHz shows a one-sided convoluted
jet, extending for $\sim$ 0\farcs5\ to the north of the quasar, and
with fainter emission extending to the east (Simon et~al. 1990).
Akujor et~al. (1994) obtained a MERLIN image of 3C~48 at 18~cm, which
slightly resolves the structure of the compact radio source.  The
radio emission extends slightly along the north-south direction, as
does the host galaxy nebulosity seen in the F555W and F814W images.
We have searched for the optical counterpart of the radio jet in our
$HST$ images, but failed to detect one; we set an upper limit of
22.2~mag~arcsec$^{-2}$ in the F814W band.

\section{B2~1425+267 (TON~202)}
\label{b2}

In the F814W band image, the host of B2~1425+267 appears to be an
elliptical galaxy, while in the F555W image, the host galaxy appears
rounder and displays an arc-like feature at an average distance from
the quasar of 1\farcs5. We have inspected dozens of WFPC2 images of
stars in our search for a stellar PSF to match the present quasar
images; none of the stellar images possess a feature like the arc seen
in B2~1425+267.  The arc could, in principle, be a spiral arm or the
debris trail of a small galaxy disrupted by the quasar host; better
images are required to reach a definitive conclusion.
The results of the one-dimensional and two-dimensional fits are
consistent with the host of B2~1425+267 being an elliptical galaxy.
In the absence of definitive evidence to the contrary, we adopt the
E-galaxy classification as being the simplest interpretation of the
available images.

There are a number of suggestions in the literature that B2~1425+267
is an interacting system.  For example, Stockton \& MacKenty (1983)
suggested that the morphology of the extended emission-line regions
around B2~1425+267 resulted from interactions between two galaxies of
similar masses.  Also, Hutchings, Crampton \& Campbell (1984) obtained
ground-based $B$ band images of B2~1425+267 and described the host
galaxy as irregular and possibly interacting with a companion.  There
are some signs in the WFPC2 images that interaction might be present;
however, it is clear that any interaction that is occurring does not
involve a galaxy of luminosity and size comparable to that of the host
galaxy.

We detect two faint resolved sources within the host galaxy, which
coincide approximately with the extended emission-line gas detected by
Stockton \& MacKenty (1983, 1987).  One source is at $\sim$ 1\farcs9
(5.8~kpc, $\rm{PA=264^\circ}$) from the center of the quasar, with
$m_{\rm F814W}=24.0$ mag.  This feature is more prominent in the F814W
image, but also appears as a higher surface brightness spot in the
F555W image, on top of the arc-like feature.  The second source has
$m_{\rm F814W} = 24.2$~mag and is located 4\farcs3 (13.2~kpc, ${\rm
PA=75^\circ}$) from the quasar.

A large number of galaxies surround B2~1425+267 in the WFPC2 images; 
three are within 10\arcsec\ (30~kpc) of the center of the host
galaxy.  Block \& Stockton (1991) detected these galaxies in their
ground-based $R$-band image and called them A, B, and C.
In the HST images galaxy~A is at 9\farcs2 ($\rm {PA=62^\circ}$),
galaxy~B at 9\farcs1 ($\rm{PA=319^\circ}$), and galaxy~C at 5\farcs4
($\rm{PA=223^\circ}$); their magnitudes in the F814W image are 21.9,
21.7 and  23.1~mag, and their colors $m_{F555W} - m_{F814W}$ are $+1.2$,
$+1.4$, and $+0.8$~mag, respectively. Galaxy~C and the 
 quasar host galaxy have similar colors, while galaxies A and B are redder.
Galaxy~C lies within the nebulosity of the
quasar host galaxy; there is no indication that the host galaxy
extends to galaxies A and B. 

Spectroscopic observations at 3\arcsec\
from the nucleus by Boroson, Person, \& Oke (1985) revealed
several emission lines in the nebulosity, including H$\alpha$, H$\beta$ and
[O~III], and some continuum emission as well. Malkan (1984)
detected a bright host for B2~1425+267, probably an
elliptical galaxy, with $M_B =  -22.1$ and $r_{\rm eff} = 17$~kpc.
Applying the Fukugita et~al. (1995) color transformation relations, 
the host galaxy measured by Malkan is 
$\gtrsim 1$~mag brighter in $V$ than our  measurements, 
and the effective radius quoted by Malkan is a factor $\sim 2$ larger than
the value given by our two-dimensional model (see Table~\ref{magtab}).

\section{3C~345 (1641$+$399)}
\label{3c345}

The bright radio source 3C~345 was optically identified by Goldsmith
\& Kinman (1965), shortly after the first quasars were discovered, and
it soon became famous for its extreme variability in almost all
wave-bands.  In the optical domain, 3C~345 is classified as a Blazar,
a subset of the radio-loud Active Galactic Nuclei (AGN) for which the
relativistic radio jet happens to point at small angles to the line of
sight (Angel \& Stockman 1980). The radio source 3C~345 exhibits
apparent superluminal motion (Schraml et~al. 1981) and high levels of
optical and radio polarization.  A one-sided radio jet extending for
several arcseconds, and a faint steep-spectrum giant halo, are also
present (Kollgaard, Wardle, \& Roberts 1989).

The $HST$ images reveal that the host galaxy of 3C~345 is a bright
elliptical (E3) that is clearly seen in the  F814W band.
No prominent peculiar morphological features are present in the two
$HST$ images. Many galaxies are detected in the quasar field,
including a bright elliptical, with $m_{\rm F814W} = 20.7$, at 26~kpc
projected separation ($\Delta\alpha = 6.8\arcsec$ and $\Delta \delta =
-1.3\arcsec$) from the quasar.  The measured redshift of this
companion galaxy, $z=0.5880$ (Ellingson \& Yee 1994), is close to the
quasar redshift ($z=0.594$).

The position angle of the major axis of the host galaxy is $\sim
60^\circ$ in the F814W image. The radio jet at several arcseconds from
the core has PA= $330^{\circ}$ (Kollgaard, Wardle, \& Roberts 1989), approximately
perpendicular to the main plane of the host galaxy.  We do not detect
an optical counterpart to the radio jet, but we cannot set strong
limits to the optical emission as one of the diffraction spike
remnants at $\rm PA=320^{\circ}$ extends for 1\farcs5 from the quasar.
The available observations show that the jet is fainter than
21.8~mag~arcsec$^{-2}$ in F814W.

\section{COMPANION GALAXIES}
\label{companion}

We have systematically examined the $HST$ images of the quasar fields for companion
galaxies projected close to the quasars.
To evaluate the statistical significance of the small projected
separations of the quasar-galaxy pairs, we constructed a sample of
``companion'' galaxies using two criteria: a maximum projected metric
displacement from the quasar, and a minimum galaxy luminosity.  

Table~\ref{comps} lists all the galaxies that have $d_{\rm proj} \le$
25 kpc and $M_{\rm F606W} \le -16.5$; these are the values used for
the Local Sample (Bahcall et~al. 1997).  The table contains the
following information: the radius (arcsec) used for the photometric
measurement, the projected distance from the quasar (in arcsec and
kpc), the apparent magnitude in the F814W band, the color index
($m_{\rm F555W} - m_{\rm F814W}$), the estimated apparent and absolute
magnitudes in F606W.  To determine the brightness of the companion
galaxies in the F606W band, we first verified, using
Figure~\ref{figgalcolors}, which morphological type corresponded to
the measured color $m_{\rm F555W} - m_{\rm F814W}$.  The color
transformation tables of Fukugita et~al. (1995) were used to obtain
$m_{\rm F606} - m_{\rm F555W}$ for each companion galaxy.

As described in \S~6 of Bahcall et~al. 1995b, to evaluate the
statistical significance of the close companions observed, we built a
comparison sample by counting galaxies in similar regions imaged at
the same time and with the same instrumental configuration as each of
the quasar fields.  We counted galaxies in circular areas that are
within 25 kpc (at the quasar redshift) of the quasar position shifted
to CCD 2 and CCD 4; the quasar images in the Local Sample are all
close to the center of CCD 3 of the WFC2.  

In the 40 fields of CCD 2 and CCD 4 that accompany the twenty quasar
images of the Local Sample, there are 8 galaxy companions that would
have satisfied the {\it a priori } criteria for inclusion in our
complete sample of close companions had they been found at the same
coordinate positions in CCD 3. For three fields, the expected number
of companions would be 0.60.
The Poisson probability of finding five or more projected close galaxy
companions as listed in Table~\ref{comps} is
%
%\begin{mathletters}
\begin{equation}
P(\geq 5, 0.6 ; {\rm 25~kpc}, M_{\rm F606W} \leq -16.5) \leq 0.0004.
\label{eq:1}
\end{equation}

If we use the WF frames (8) that accompany the three RLQs to draw a
control sample, we find 6 galaxies that satisfy the selection criteria.
For three fields, the expected number of companions would be 2.3. 
The Poisson probability of finding five or more projected close galaxy
companions is
%
%\begin{mathletters}
\begin{equation}
P(\geq 5, 2.3 ; {\rm 25~kpc}, M_{\rm F606W} \leq -16.5) \leq 0.05.
\label{eq:2}
\end{equation}

We conclude that the presence of the projected close companions around
the three RLQs is statistically significant. A similar result was
found by Bahcall (1995b,1997) and Fisher et~al. (1996) for the Local
Sample.

Ellingson \& Yee (1994) measured the redshifts of 31 galaxies in the
field of 3C~345.  Eight of their galaxies have velocities within
3000~${\rm km~s^{-1}}$ of the quasar. Five of those galaxies are included
 in the $HST$ images, although only one of them is close
enough to appear in the field of view shown in Figures~\ref{figqso}
and \ref{figqsosub}.

The five galaxies in the HST images known to be at the same distance
as 3C~345 have measured colors $m_{\rm F555W} - m_{\rm F814W}$ that
are consistent to the colors expected for normal galaxies of similar
morphology and redshift.

\section{DISCUSSION}
\label{discuss}

The main results of this study are: 1) the three RLQs studied have
bright (several $L^*$) host galaxies; 2) the hosts of RLQs are, on
average, about 1~mag more luminous than the hosts of RQQs; and 3) the
host colors $m_{\rm F555W} - m_{\rm F814W}$ are substantially bluer
than expected for galaxies of similar morphological types at the same
redshifts as the quasars.

\subsection{Nature of the Hosts}
\label{subnature}

Ground-based observations have produced contradictory evidence
regarding the stellar populations of the quasar hosts galaxies.
Orndhal et~al. (1996) report that quasar host galaxies have $V-R$
colors typical of late-type spirals, whereas Kukula et~al. (1996),
concluded from near-infrared observations that the non-interacting
hosts of RLQs are systems with old, red stellar populations.  The
ability of $HST$ to address this question suggests that one should
further investigate the colors of the hosts of the low-redshift
quasars to determine whether 1)~the host galaxies of luminous quasars
are generally bluer than ``normal'' galaxies of similar morphological type,
2)~host galaxies of RLQs are bluer than the hosts of RQQs, or~3)~we
are witnessing an extraordinary evolutionary effect in the colors of
host galaxies.

The $HST$ images of 3C~48 reveal tidal tails extending for $\sim
35$~kpc, confirming previous suggestions of a strong gravitational
interaction in this system.  Evidence for a massive host was presented
earlier from ground-based imaging and spectroscopy (Boroson \& Oke
1982; Malkan 1984; Stockton \& Ridgway 1991), observations of CO
(Scoville et~al. 1993) and far-infrared emission (Neugebauer et
al. 1985). The host galaxy of 3C~48 has $M_V = -22.7$~mag, brighter
than a typical brightest cluster galaxy ($M_V~\approx~-22.0$, Hoessel \&
Schneider 1985; Postman \& Lauer 1995) for $\Omega_0 = 1.0$ and $H_0 =
100$ km s$^{-1}$Mpc$^{-1}$, and brighter than any host in the 20
member Local Sample.

The hosts of B2~1425+267 and 3C~345 both have light distributions
similar to luminous elliptical galaxies. However, for B2~1425+267, the
presence of clumps and a faint galaxy lying within the host galaxy 
 may indicate ongoing gravitational interaction.

All nine RLQs we have studied (either in the Local Sample or in this
paper) observed with $HST$ have light distributions consistent with
their hosts  being elliptical galaxies, or occurring in
interacting systems; in some cases there is evidence for both.

\subsection{Hosts of RLQs are brighter than hosts of RQQs}
\label{subhosts}

Several groups have reported on a positive correlation between the
luminosity of the active nucleus and that of the host galaxy (McLeod
\& Rieke 1994b, 1995a; Hooper et~al. 1997; Kotilainen et~al. 1998).
McLeod \& Rieke (1995a, 1996) suggested that for low-redshift quasars
brighter than ${M_B \lesssim -22.4}$, there is a minimum mass for the
host galaxy that increases with increasing nuclear power, implying
that a more luminous host galaxy is required to fuel a more luminous
quasar.  Laor (1998) found a correlation between the luminosity of the
quasar host galaxies in the Local Sample, and the inferred (from
H$\beta$) black hole mass. Laor points out that the correlation is
similar to the black hole mass versus bulge luminosity correlation
found for normal galaxies (Magorrian et~al. 1998),
suggesting that quasars reside in normal galaxies.

Figure~\ref{figmv} shows the two-dimensional model absolute $V$
magnitudes (from Table~\ref{magtab}) of the host galaxies versus the
absolute $V$ magnitudes of the quasars (Table~\ref{journal}) in the
present study and in the Local Sample.  The best linear fit to all the
data yields:
\begin{equation}
 M_V({\rm host}) = 0.52 \ M_V({\rm QSO}) -8.67.
\label{eq:3}
\end{equation}
The linear correlation coefficient (Pearson's r) indicates only a 
$1.3\%$ probability for the optical absolute magnitudes of the quasar
and the host to be uncorrelated.

The two most striking aspects of the Figure~\ref{figmv} visible by eye
are: 1)~there is essentially no correlation of $M_V({\rm host})$ and
$M_V({\rm QSO})$ if we consider the RQQs
and RQLs separately (probability of no correlation higher than $\sim
50\%$), and 2)~the RLQs have host galaxies that are about a magnitude
brighter than the host galaxies of RQQs.

Several previous studies have compared the luminosities of the host
galaxies of RQQs and RLQs, but a consensus has not
been achieved.  There are suggestions that the hosts of
RLQs are more luminous than the hosts of RQQs (Hutchings, Janson \& Neff 1989;
V\'eron-Cetty \& Woltjer 1990; McLeod \& Rieke 1995a, Bahcall et
al. 1997) and also that there is no significant difference between the
host luminosities (Dunlop et~al. 1993, Orndahl et~al. 1996,
Taylor et~al. 1996, Hooper et~al. 1997).

For the 23 members of our $HST$ sample (20 in the Local Sample and 3
RLQs discussed in this paper), the hosts of the RLQs are on average 1~mag brighter
than the hosts of the RQQs.
The average best-fit 2-D model magnitudes for the hosts of the 14
RQQs and 9 RLQs are:
\begin{equation}
\langle M_V \rangle _{\rm host,RQQ} = -20.6 \pm 0.2~{\rm
mag},
\label{eq:4}
\end{equation}
\begin{equation}
\langle M_V \rangle _{\rm host,RLQ} = -21.7 \pm 0.2~{\rm mag}.
\label{eq:5}
\end{equation}
where the error given is the error of the mean (the standard deviation
is $\sqrt N$ times larger).

We applied the Student's t-test to measure the significance of the
difference of means.  The probability that the host galaxies of the
RLQs and RQQs have the same mean absolute $V$ magnitude is $1.3\times
10^{-4}$.  The Kolmogorov-Smirnov test give a probability of
$9\times10^{-4}$ for the distributions of the absolute magnitudes of
the host galaxies of the radio-loud and radio-quiet quasars to be the
same.

One must view these statistical tests with caution, however, as a
glance at Figure~\ref{figmv} shows that the distributions of the
optical luminosities of the RQQs and RLQs are significantly different.
The four most optically luminous quasars in the entire sample are
RLQs, and the average luminosity of the classes differs by nearly  a
magnitude
\begin{equation}
\langle M_V \rangle _{\rm RQQ} = -23.3 \pm 0.1~{\rm mag} \ \ \ \ \
({\rm 14 \ objects}),
\label{eq:6}
\end{equation}
\begin{equation}
\langle M_V \rangle _{\rm RLQ} = -24.2 \pm 0.3~{\rm mag}\ \ \ \ \
({\rm 9 \ objects}).
\label{eq:7}
\end{equation}
If the luminosity of the host depended on the optical luminosity of
the quasar, our
sample would indicate a RQQ/RLQ difference even if the host galaxy
luminosity was independent of quasar radio properties.

We selected a subsample of quasars in which the RLQs and the RQQs have
the same absolute magnitude range in order to limit the comparison to quasars
with similar optical properties.  In the range $-24.5 \leq M_V(\rm
QSO) \leq -23.2$, there are 6 RQQs and 7 RLQs; both subsamples of
quasars have the same average absolute $V$ magnitudes ($\langle
M_V({\rm QSO})\rangle = -23.8$~mag).  For those two subsamples the
average best-fit 2-D model magnitudes of the host galaxies (and error
of the mean) are:
\begin{equation}
\langle M_V \rangle _{\rm host,RQQ} = -20.6 \pm 0.2~{\rm mag},
\label{eq:8}
\end{equation}
\begin{equation}
\langle M_V \rangle _{\rm host,RLQ} = -21.7 \pm 0.3~{\rm mag}.
\label{eq:9}
\end{equation}
The host galaxies of RLQs  in this group are on average 1.1~mag (3.1
$\sigma$) brighter than the hosts of the RQQs, similar to the result
obtained for the whole sample.

Could the difference be an artifact of the galaxy model used in the
fitting?  This is a concern because of the extrapolations required to
calculate the total luminosity; the very center of the galaxy is
masked by the quasar, and one can only fit out to a few (at best)
effective radii or scale lengths. Since we are restricted to data at
intermediate radii, a de Vaucouleurs model fit produces 
a total
magnitude that is systematically brighter than the value obtained with
an exponential disk, because the de Vaucouleurs model rises more
steeply in the very center and falls more slowly at large radii than
does a disk model.  For the 23 quasars in our sample, the
two-dimensional GdV model gives magnitudes for the host that are on
average 0.6~mag brighter than the exponential disk estimates.

For the limited 13 member sample with $-24.5 \leq M_V({\rm QSO}) \leq
-23.2$, we fit all the hosts with a two-dimensional GdV galaxy model,
and also with an exponential disk model. In both cases, the hosts of
the RLQs are on average 0.8~mag brighter than the hosts of the RQQs
(difference higher than 3$\sigma$), implying that the difference in
the host luminosities of RLQs and RQQs is real.

Figure~\ref{figradio} shows the radio luminosity of the quasars at
6~cm, from the V\'eron-Cetty \& V\'eron (1996) catalog, versus the
two-dimensional model absolute visual magnitudes of the host
galaxies. Figure~\ref{figradio} includes 9 RLQs and 8 RQQs for
which V\'eron-Cetty \& V\'eron (1996) list a detected emission at
6~cm.  This figure shows that the strongest radio sources tend to
occur in more luminous host galaxies. As seen in Figure~\ref{figmv},
there is no trend for correlation inside the two radio classes.
The trend shown in  Figure~\ref{figradio} can be represented by a
linear regression:
\begin{equation}
 M_V({\rm host}) = -0.36 \ \log L_{\rm 6cm} -12.33.
\label{eq:10}
\end{equation}
The linear correlation coefficient indicates that there is less than a
0.1\% probability of the luminosity of the host and the
quasar radio power being uncorrelated.  

We conclude that the difference in the luminosities of the host
galaxies of the RLQs and RQQs is significant and robust. The apparent
correlation between the optical luminosities of the host and the
quasar (Figure~\ref{figmv}) is mainly due to the RLQs occurring in
brighter galaxies than the RQQs.  The results suggest that the radio
emission of the quasars is connected with the magnitude/mass of the
host galaxy, indicating that only galaxies brighter than a certain
limit would be able to harbor a radio-loud quasar.

\acknowledgments 

We dedicate this paper to the memory of our colleague and friend
Jerome Kristian, an outstanding scientist who played a key role in the
development of this field of research.  Kristian was a member of the
HST-WFPC team that obtained the high quality data shown in this paper.
We thank the anonymous referee whose comments improved the
presentation of the paper. We are grateful to Tal Alexander, Ari Laor,
and Insu Yi for their comments and suggestions, and to Maggie Best for
her help with the tables.  David Saxe and Robert Deverall wrote much
of the code used in the analysis presented in this paper.  We would
like to thank Digital Equipment Corporation for providing the DEC4000
AXP Model 610 system used for the computationally intensive parts of
this project.  This work was supported in part by NASA contract
NAG5-3259, NASA grant number NAG-5-7047 and grant number GO-5343 from
the Space Telescope Science Institute, which is operated by the
Association of Universities for Research in Astronomy, Incorporated,
under NASA contract NAS5-26555.  We have used the NASA/IPAC
Extragalactic Database (NED), operated by the Jet Propulsion
Laboratory, Caltech, under contract with NASA, and NASA's Astrophysics
Data System Abstract Service (ADS).

\newpage

\vfil\eject

\newpage

\centerline{\bf FIGURE CAPTIONS}

\begin{figure}[h]
\caption[]{\baselineskip=14pt 
This figure shows 
the $14\arcsec \times 14\arcsec$\ PC images of 3C~48 and 3C~345 and
$30\arcsec \times 30\arcsec$\ WF3 images of B2~1425+267, taken with
filters F555W (left-hand panels) and F814W (right-hand panels).  The
images shown are the average of two 1400~s exposures in F555W and two
1700~s exposures in F814W (see Table~\ref{journal}). Cosmic ray
subtraction and pipeline STScI flatfielding are the only processing
performed on the data. The images have been rotated
in order to have north on top and east to the left.
\protect\label{figqso}}
\caption[]{\baselineskip=14pt  
Same display as shown in 
Figure~\ref{figqso}, but with a best-fit PSF subtracted from the
data. An arrow in the F555W image of 3C~48 indicates a ring-feature
which may be an optical artifact.
\protect\label{figqsosub}}
\caption[]{\baselineskip=14pt 
The predicted $m_{\rm F555W} - m_{\rm F814W}$ color index for 
different galaxy morphological types as a function of the redshift,
from Fukugita et~al. (1995). The color index for the three quasar host
galaxies (from 2-D fits listed in Table~\ref{magtab}) are indicated.
\protect\label{figgalcolors}}
\caption[]{\baselineskip=14pt 
Bright knots observed in the broad-band
and [O~III]$\lambda$ 5007 images of the 3C~48 host galaxy.  The
upper panels show the F555W (right) and F814W (left) images.
The lower panels show the [O~III] 
image taken with the linear ramp filter FR680N18.  The
lower-left panel has a higher contrast stretch to show the bright
[O~III] knot at 0\farcs8 to the south of the quasar.  The knots are
identified by Roman letters and listed in Table~\ref{3c48knots}.  Each
image is 24\arcsec\ on a side with north at the top and east to the
left.  The [O~III] image was taken with the WF2 while the F555W and
F814W images were taken with the PC.  \protect\label{fig3c48knots}}
\caption[]{\baselineskip=14pt 
Absolute $V$ magnitude of the host galaxies versus the absolute
$V$ magnitude of the quasars. The plotted values include the 
20 Local Sample  quasars and the three radio-loud quasars discussed in
this paper.
The radio-quiet quasars are represented
by filled circles, while stars represent the radio-loud quasars.
\protect\label{figmv}}
\caption[]{\baselineskip=14pt
Absolute $V$ magnitude of the host galaxies versus the radio
luminosity of the quasars at 6~cm. The flux densities at 6~cm were taken
from V\'eron-Cetty \& V\'eron (1996). Quasars not detected in the
radio are indicated by an arrow. The sample and symbols are the same
as in Figure~\ref{figmv}.
\protect\label{figradio}}
\end{figure}

%% TABLES

\vfil\eject

%Table 1 
%\begin{deluxetable}
%{l@{\hspace{2pt}}c@{\hspace{3pt}}l@{\hspace{2pt}}l@{\hspace{0pt}}
%l@{\hspace{-5pt}}c@{\hspace{-5pt}}c@{\hspace{1pt}}
%c@{\hspace{1pt}}l@{\hspace{3pt}}c}
\begin{deluxetable}{lclcccclc}
\footnotesize
\tablewidth{7.2in}
\tablecaption{Journal of Observations\label{journal}}
\tablehead{
&&&&\colhead{Sky Level}&&&\colhead{kpc$^{\rm c}$}&\\
\colhead{Quasar}&\colhead{Date}&\colhead{Filter}
&\colhead{CCD}&
\colhead{(${\rm e^{-} pix^{-1} s^{-1}}$)}&
\colhead{$z^{\rm b}$}&\colhead{$V_{\rm QSO}^{\rm b}$}&\colhead{${\rm arcsec^{-1}}$}
&\colhead{$M_{V(\rm QSO)}^{\rm c,d}$}}
\startdata
3C~48       & 21 Jan 95& F555W & PC1& 0.015&  0.367& 16.20& 3.08~(6.62)& $-$24.0~($-$25.7) \nl
            & 21 Jan 95& F814W & PC1& 0.014&       & & & \nl                  
            & 30 Jan 96& FR680N18$^{\rm a}$& WF2&$<$0.001& & & & \nl                  
B2~1425+267 & 02 Dec 94& F555W & WF3& 0.085&  0.366& 15.68& 3.07~(6.61)& $-$24.5~($-$26.2) \nl
            & 02 Dec 94& F814W & WF3& 0.085& & & & \nl                  
3C~345      & 20 Oct 94& F555W & PC1& 0.009&  0.594& 15.96& 3.79~(8.51)& $-$25.3~($-$27.0) \nl
            & 20 Oct 94& F814W & PC1& 0.008& & & & \nl                  
\enddata
\tablenotetext{a} { Linear ramp filter centered at 6845~\AA,
which covers the [O~III]$\lambda$ 5007~\AA\ emission at the quasar redshift.}
\tablenotetext{b} { From V\'eron-Cetty \& V\'eron (1996).}
\tablenotetext{c} { Computed for $\Omega_0 = 1.0$ and $H_0 = 100$ km s$^{-1}$Mpc$^{-1}$
($\Omega_0 = 0.2$ and $H_0 = 50$ km s$^{-1}$Mpc$^{-1}$). }
\tablenotetext{d} { We assumed that the spectrum
of quasars can be described by a power-law of the form $f_\nu \propto
\nu^{-0.5}$.}
\end{deluxetable}

%Table 
\begin{deluxetable}{lrrcclc}
\footnotesize
\tablewidth{6.8in}
\tablecaption{Size and Morphology of Host Galaxies\label{size}}
\tablehead{
\colhead{Quasar}&\colhead{Filter}&\multicolumn{2}{l}{Major\ Diameter}
&\colhead{Isophote}&\multicolumn{1}{l}{Morphology}\\
&&\colhead{\arcsec}&\colhead{kpc$^{\rm a}$}
&\colhead{mag arcsec$^{-2}$}&\colhead{(apparent)}}
\startdata
3C~48         & F555W &  8.5& 26~(56)& 23.3& Interacting    &\nl
3C~48         & F814W &  9.8& 30~(65)& 23.2& Interacting    &\nl
B2~1425+267   & F555W & 11.2& 34~(74)& 25.2&  E1 &\nl
B2~1425+267   & F814W & 14.2& 44~(94)& 25.1&  E3 &\nl
3C~345        & F555W &  4.5& 17~(38)& 23.8&  E2 &\nl
3C~345        & F814W &  6.5& 25~(55)& 23.6&  E3 &\nl
\enddata
\tablenotetext{a} { Computed for $\Omega_0 = 1.0$ and $H_0 = 100$ km
s$^{-1\ }$Mpc$^{-1}$ ($\Omega_0 = 0.2$ and $H_0 = 50$ km
s$^{-1}$Mpc$^{-1}$).  }
\end{deluxetable}

%Table 3

\begin{table}
\small
\caption{Magnitudes of the Quasar Host Galaxies}
\bigskip
\centerline{Aperture Photometry\label{magtab}}
\begin{tabular*}{7.0in}{@{\extracolsep{\fill}}lcccccc}
\noalign{\smallskip}
\hline\hline
\noalign{\smallskip}
Quasar&inner&outer&$m_{\rm F555W}$&$m_{\rm F555W}~ -$&$M_{V}^{\rm 
a}$&$M_{I}^{\rm a}$\\
&radius&radius&&$m_{\rm F814W}$&&\\
&(\arcsec)&(\arcsec)&&\\
\noalign{\smallskip}
\hline
\noalign{\smallskip}
3C~48          &   0.3&  5.5&  18.1& \llap{$+$}0.7& $-$22.3~($-$23.9)&  
$-$23.2~($-$24.8) \\
B2~1425+267    &   0.6&  4.0&  19.1& \llap{$+$}0.7& $-$21.3~($-$22.9)&  
$-$22.2~($-$23.8) \\
3C~345         &   0.3&  2.3&  20.0& \llap{$+$}1.3& $-$21.0~($-$22.8)&  
$-$22.8~($-$24.6) \\
\noalign{\smallskip}
\hline
\end{tabular*}
\vglue.2truein
%Table 4
\small
\centerline{Analytic Galaxy Model Fits}
\begin{tabular*}{7.0in}{lcccccccc}
\noalign{\smallskip}
\hline\hline
\noalign{\smallskip}
Quasar&$m_{\rm F555W}$&$r(\arcsec)^{\rm b}$&$m_{\rm F814W}$&
$r(\arcsec)^{\rm b}$&$m_{\rm F555W}~ - $&Best&$M_{V}^{\rm a}$&
$M_{I}^{\rm a}$\\
&&&&&${m_{\rm F814W}}$&Model\\
\noalign{\smallskip}
\hline
\noalign{\smallskip\smallskip\centerline{One-Dimensional}\smallskip}
\hline
\noalign{\smallskip}
3C~48          & 17.9& 1.2 & 17.1& 0.9 & \llap{$+$}0.8 & GdV& 
$-$22.5~($-$24.1)& $-$23.5~($-$25.1) \\
B2~1425+267    & 18.4& 1.5 & 17.8& 1.5 & \llap{$+$}0.6 & GdV& 
$-$22.0~($-$23.6)& $-$22.8~($-$24.4) \\
3C~345         & 19.4& 0.3 & 18.1& 0.8 & \llap{$+$}1.3 & GdV& 
$-$21.6~($-$23.4)& $-$23.4~($-$25.2) \\
\noalign{\smallskip}
\hline
\noalign{\smallskip}
\noalign{\smallskip\centerline{Two-Dimensional}\smallskip}
\hline
\noalign{\smallskip}
3C~48          & 17.7& 2.5 & 17.2& 1.8 & \llap{$+$}0.5 & GdV & 
$-$22.7~($-$24.3) & $-$23.4~($-$25.0) \\
B2~1425+267    & 18.3& 2.0 & 17.6& 2.7 & \llap{$+$}0.7 & GdV & 
$-$22.1~($-$23.7) & $-$23.0~($-$24.6) \\
3C~345         & 19.6& 0.5 & 18.2& 0.6 & \llap{$+$}1.4 & GdV & 
$-$21.4~($-$23.2) & $-$23.3~($-$25.1) \\
\noalign{\smallskip}
\hline
\noalign{\bigskip}
\noalign{\vbox{\hsize=6.8in\noindent
${\rm ^a}$\ Computed for $\Omega_0 = 1.0$ and $H_0 = 100$ km
s$^{-1\ }$Mpc$^{-1}$ ($\Omega_0 = 0.2$ and $H_0 = 50$ km
s$^{-1}$Mpc$^{-1}$).  
$k$-corrections from Fukugita et~al. (1995)}
${\rm ^b}$\ Effective radius.\hfil\break
}
\end{tabular*}
\end{table}

%TABLE 4: Bright knots in 3C 48

\begin{deluxetable}{cccccccc}
\footnotesize
\tablewidth{15cm}
\tablecaption{Bright Knots observed in the Broad-Band
and [O~III] Images of the Host Galaxy of 3C~48\
\label{3c48knots}}
\tablehead{
\colhead{$d$}&\colhead{$\Delta\alpha$}&\colhead{$\Delta\delta$}
&\colhead{Aperture$^{1}$}&\colhead{$m_{\rm F814W}$}&
\colhead{$m_{\rm F555W}$}&
\colhead{[O~III] flux$^{2}$}&\colhead{Region$^{3}$}\\
\colhead{($\arcsec$)}&\colhead{($\arcsec$)}&\colhead{($\arcsec$)}
&\colhead{Radius}&\colhead{(mag)}&\colhead{(mag)}&
\colhead{(${\rm 10^{-16}~erg~cm^{-2}~s^{-1}}$)}\\
&&&\colhead{($\arcsec$)}&&
}
\startdata 
0.8 &          0.2 & \llap{$-$}0.8 & $-$& $-$ & $-$ & 23.6 & a\nl
1.1 &          1.0 &           0.5 & 0.2& 21.8& 22.6&  $-$ & b$^{4}$\nl
1.4 &          0.3 &           1.4 & 0.3& 22.8& 24.0&  2.9 & c\nl
1.8 &\llap{$-$}0.2 &           1.8 & $-$& $-$ & $-$ &  2.3 & d\nl
2.1 &          0.1 &           2.1 & $-$& $-$ & $-$ &  2.2 & e\nl
2.1 &          1.2 & \llap{$-$}1.7 & 0.4& 23.5& $-$ &  $-$ & f\nl 
2.2 &\llap{$-$}0.8 &           2.0 & 0.2& 25.6& 23.5&  $-$ & g\nl 
2.4 &          0.9 & \llap{$-$}2.2 & 0.1& 24.8& 25.1&  $-$ & h\nl 
2.5 &          2.5 &           0.4 & $-$& $-$ & 26.2&  2.9 & i\nl
2.7 &          0.6 & \llap{$-$}2.6 & 0.1& 25.1& 26.7&  $-$ & j\nl 
2.8 &\llap{$-$}0.5 &           2.8 & 0.2& 25.9& 25.5&  $-$ & k\nl 
3.6 &          0.6 &           3.6 & 0.3& 24.4& 24.6&  8.1 & l\nl 
3.7 &\llap{$-$}1.6 &           3.3 & 0.1& 25.6& $-$ &  $-$ & m\nl
4.1 &\llap{$-$}1.4 &           3.8 & 0.2& 25.4& 25.9&  $-$ & n\nl 
4.1 &          0.3 &           4.1 & 0.3& 24.6& 24.5&  7.1 & o\nl 
4.6 &\llap{$-$}2.1 &           4.1 & 0.2& 24.7& 25.0&  $-$ & p\nl 
4.7 &\llap{$-$}0.6 &           4.7 & 0.3& 25.4& 25.7&  2.8 & q\nl
\enddata  
\tablenotetext{1} { Broad-band images.}
\tablenotetext{2} { Measured within a 0.3\arcsec\ radius.}
\tablenotetext{3} { Nomenclature from Figure~\ref{fig3c48knots}.}
\tablenotetext{4} { This knot is probably part of an image artifact (see
\S~\ref{sub3c48broad}).}
\end{deluxetable}

% TABLE  - COMPANIONS 
\begin{deluxetable}{lccl@{\extracolsep{17pt}}ccccc}
%\begin{deluxetable}{lcccccccc}
\footnotesize
\tablewidth{6.8in}
\tablecaption{Galaxy Companions with $M_{\rm F606W}$ brighter than
$-16.5$~mag  within 25~kpc ($\sim 55~{\rm kpc})^{\rm a}$ of the Quasar  
\label{comps}}
\tablehead{
\colhead{Quasar}&\colhead{Number}&\colhead{Aperture}
&\multicolumn{2}{c}{Distances}&\colhead{$m_{\rm F814W}$}&{$m_{\rm
F555W}~ -$}&\colhead{$m_{\rm F606W}^{\rm b}$}&\colhead{$M_{\rm F606W}$}\\
&\colhead{ of }&\colhead{Radius}&\colhead{($\arcsec$)}&\colhead{kpc$^{\rm a}$}
&&\colhead{$m_{\rm F814W}$}\\
&\colhead{Companions}&\colhead{($\arcsec$)}&&&&}
\startdata 
3C~48        &1& 0.7&  7.0& 21.5~(46.3)& 23.4& \llap{$+$}0.9& 23.9& \llap{$-$}16.5 \nl

B2~1425+267  &1& 1.0&  5.4& 16.6~(35.7)& 23.1& \llap{$+$}0.8& 23.4& \llap{$-$}17.0 \nl

3C~345       &3& 0.3&  2.8& 10.6~(23.8)& 23.9& \llap{$+$}1.1& 24.6& \llap{$-$}16.9\nl
             & & 0.2&  3.6& 13.7~(30.6)& 23.3& \llap{$+$}2.0& 24.7& \llap{$-$}16.8\nl
             & & 0.4&  5.1& 19.3~(43.4)& 23.0& \llap{$+$}0.6& 23.3& \llap{$-$}18.2\nl
\enddata
\tablenotetext{a} { Computed for $\Omega_0 = 1.0$ and $H_0 = 100$ km
s$^{-1\ }$Mpc$^{-1}$ ($\Omega_0 = 0.2$ and $H_0 = 50$ km
s$^{-1}$Mpc$^{-1}$).}
\tablenotetext{b} { Color transformation from Fukugita et
al. (1995). See \S~\ref{companion} for details.}
\end{deluxetable}

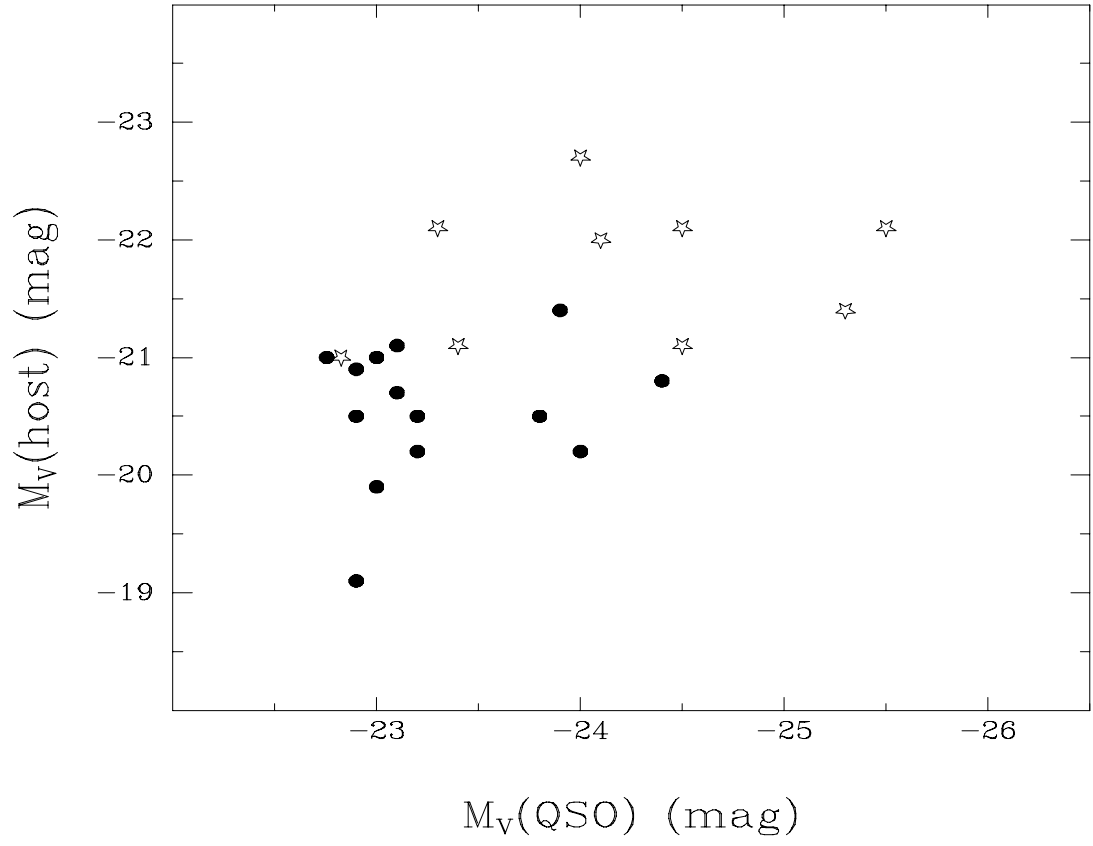
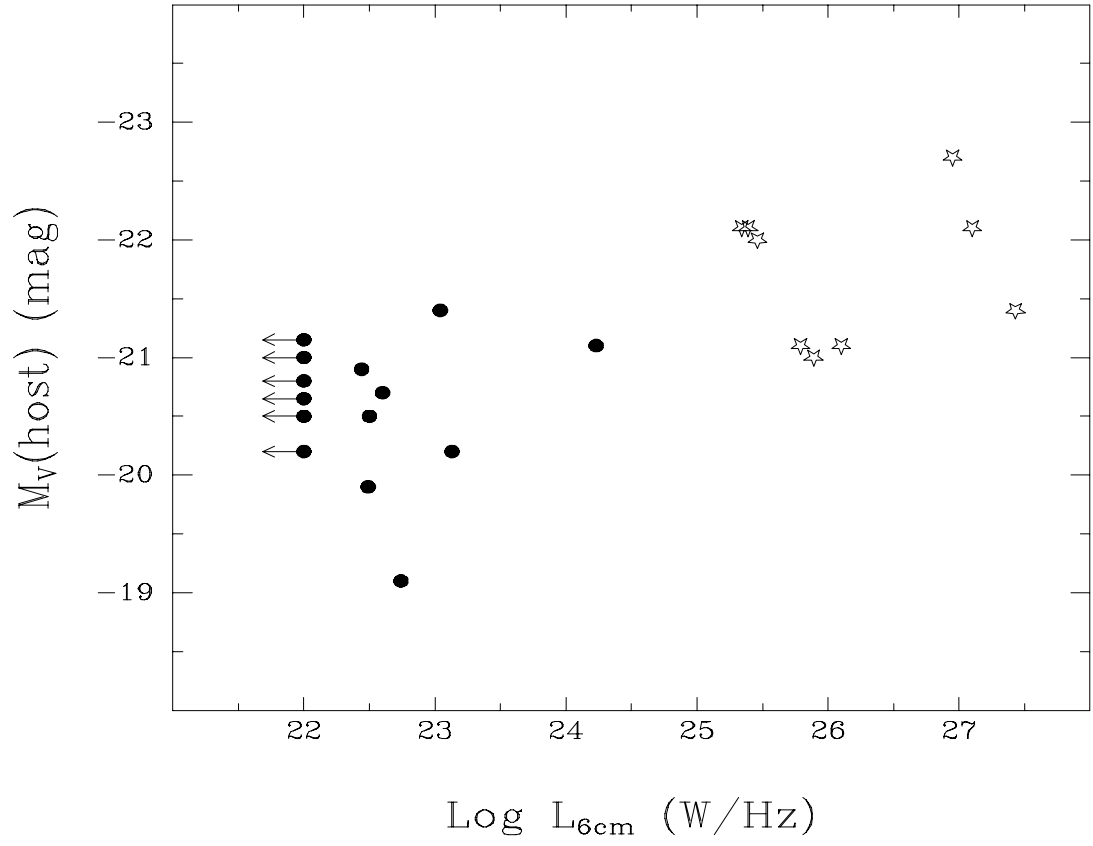
\end{document}